\documentclass[vecphys]{svmult}
\usepackage{makeidx}         
\usepackage{graphicx}        
\usepackage{multicol}        
\usepackage[bottom]{footmisc}
\makeindex             
\begin{document}
\title*{Evolutionary self-consistent models of HII galaxies}
\author{M. Mart\'{\i}n Manj\'{o}n\inst{1}\ ,
M. Moll\'{a}\inst{2}\and A.I. D\'{\i}az\inst{3}}
\institute{Universidad Aut\'{o}noma de Madrid
\texttt{mariluz.martin@uam.es} \and CIEMAT
\texttt{mercedes.molla@ciemat.es}\and Universidad Aut\'{o}noma de
Madrid \texttt{angeles.diaz@uam.es}}
%
%
\maketitle
\index{Author1}
\index{Author2}
\index{Author3}
We have studied the viability of new theoretical models which
combine a chemical evolution code, an evolutionary synthesis code
and a photoionization code, to understand the star formation and
evolution of  H{\sc ii} galaxies.  The emission lines observed in
H{\sc ii} galaxies are reproduced by meas of the photoionization
code CLOUDY,  using as ionizing spectrum the spectral energy
distribution of the modeled H{\sc ii} galaxy, which, in turn, is
calculated according to a  Star Formation History (SFH) and a
metallicity evolution given by a chemical evolution model. Our
technique reproduces the observed diagnostic diagrams and equivalent
width-color correlations for local H{\sc ii} galaxies.
\section{Introduction}
\label{sec:1}
 Blue Compact Galaxies (BCG), specially H{\sc ii} galaxies, are gas rich
dwarf galaxies which, according to the observed strong and narrow
emission lines that dominate their optical spectra, are experiencing
intense star formation processes in very reduced volumes. These
emission lines are produced by massive stars that ionize the
surrounding gas which implies the existence of a very young stellar
population. On the other hand, the analysis of their emission line
spectra shows that H{\sc ii} galaxies are very metal poor systems.
These two facts taken together lead to the idea that actually these
systems were very young objects experiencing their first burst of
star formation \cite{thuan95}. However, recent photometric
observations indicate that they host stellar populations which are
at least $10^{7}-10^{8}$ years old, reaching even 1 Gyr is some
cases \cite{leg00,tols03,cai03,thuan05}. At present, there seems to
be a general agreement about the presence of underlying populations
with ages of several Gyrs in BCGs and/or H{\sc ii} galaxies, that
is, about the no existence of genuinely young galaxies in our local
universe, but there is no consensus about how these stars formed.

There are some  works that combine stellar synthesis techniques with
photo-ionization codes (\cite{garc95}). The synthesis models assume
that a certain mass of stars is created in a single burst of
formation, what implies that all stars are similarly old and have
similar metal composition. These generations of stars are called
{\sl Single Stellar Populations} (SSP). These models obtain as a
result the spectral energy distribution of the population (SED) as
the sum of the spectra of each individual star. After computing this
SED, it is included in a photo-ionization code as the ionization
source to compute the emission line spectrum. The addition of the
chemical evolution allows to follow the metallicity of successive
stellar generations. To our knowledge, a combination of chemical
evolution, evolutionary population synthesis and photo-ionization
models has not yet been attempted to analyze the BCG or H{\sc ii}
galaxies. The potential of the method resides in the simultaneous
use of the available information about the SFH, the photometric
evolution and ionizing properties of the object: chemical
abundances, SED or colors and emission line intensities. In this
contribution we show that some of these models can adequately
reproduce these constraints.

\section{The models}
\label{sec:2}
 The aim of this work is to check the
possibility to face the problem of analyzing the H{\sc ii} galaxies
data through the combination of the three techniques mentioned above
in a self-consistent way, that is using the same assumptions about
stellar evolution, stellar atmospheres and nucleosynthesis, and
using a realistic age-metallicity relation. The procedure is to
first compute chemical evolution models (\cite{fer92,mol96,mol99})
which reproduce the observed chemical abundances and derive the SFH
and the age-metallicity relation that is then in the computation, by
means of evolutionary synthesis models, of the corresponding SED's
(\cite{garc95,mol00}). Finally, we use the photo-ionization code
CLOUDY (Ferland, 1991) using as inputs the calculated abundances and
the synthesized spectra, to obtain the emission line intensities of
the ionized gas. The combination of the results from
spectrophotometric and photo-ionization models is used to calculate
the line equivalent widths. The self-consistency and evolutionary
nature of the models permit to calculate any  quantity as a function
of time, allowing their application to objects at different
redshift.

We have run models starting with a total mass of gas of
100$\cdot$10$^{6}$ M$\odot$ concentrated inside a spherical volume
radius of 500 pc. This is the characteristic size of starbursts in
HII galaxies measured on CCD images \cite{tel97}. We have assumed
star formation to proceed in 10 consecutive bursts along 13.2 Gyrs,
one every 1.3 Gyr. In every star formation episode, gas is consumed
at a rate that depends on the available gas,  being a decreasing
function of time through a parameter that defines the star formation
efficiency. In the models presented here each burst starts with an
initial star formation efficiency  $\epsilon$ and with successive
bursts having initial efficiencies reduced by a factor $n$ (number
of the burst). We call these: {\it attenuated burst models}. We have
run models with three different initial star formation efficiencies,
$\epsilon$: 0.64 (model M1), $\epsilon$ = 0.33 (M2) and $\epsilon$ =
0.10 (M3). The code calculates the chemical abundances of 15
different elements as a a function of time at different ages in
steps of $\Delta$t= 0.5 Myrs, from $t=0$, to $t= 13.2 Gyrs$ that
would be the age of the assumed galaxy. Once the chemical evolution
is calculated, the spectrophotometric properties corresponding to
the whole stellar population are obtained, including the ionizing
continuum of the stellar burst last formed. This continuum is
introduced in the photo-ionization code together with its calculated
abundances to calculate the properties of the ionized gas which is
assumed to be at a distance R = 500 pc and to be of constant
density.

\section{Results.}
\subsection{Star formation rate (SFR).}
The SFR is one of the main differences among models since it leads
the behavior of all the other quantities. In Fig.1 we can see the
SFH for our models. In all of them the first burst is intense, while
the successive bursts are weaker because the amount of gas available
for the next burst has decreased, in spite of the gas ejected by the
evolving massive stars, and due to the attenuation of the efficiency
included in the inputs.
\begin{figure}
\centering
\includegraphics[height=4cm]{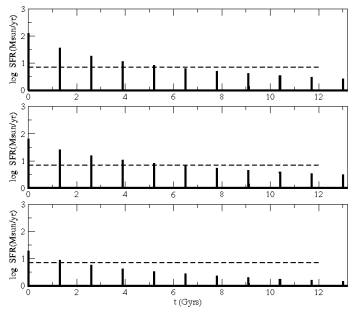}
%
%
\caption{Star formation rate of model M1 (upper panel), model M2 (middle one)
and models M3 (bottom panel)}
\label{fig:sfr}       
\end{figure}
\cite{hoy04} have estimated the SFRs for a sample of 39 local H{\sc
ii} galaxies of different morphologies from \cite{tel97}, being in
the range from 0.5 to 7 M$\odot$/yr. Our models cover this range,
whose upper bound is given by the broken line in the figure, during
the five firsts bursts in M1 and M2 and after the second burst in
M3.
\subsection{Evolution of the oxygen abundance.}
The value of the oxygen abundance, which defines the metal content
of stars and gas, increase with time due to the ejection of
processed gas from massive stars.  Observational data from
\cite{ter91} and \cite{hoy06} show that most  H{\sc ii} galaxies
have gas oxygen abundances between $7.5 < 12+log(O/H) < 8.5$. We
show these limits in Fig.2 as dashed lines. In this figure it can be
seen that M1 reaches values over the upper limit of the observations
right since the occurrence of the first burst. The efficiency of
this first bursts is very high, forming a great number of stars, and
then the metallicity grows rapidly. M2 and M3 show oxygen abundances
in this range during the whole evolution  due to the progressive
attenuation of the bursts that makes the star formation smoother so
that the oxygen abundance does not reach values as high as in M1.
\begin{figure}
\centering
\includegraphics[height=4cm]{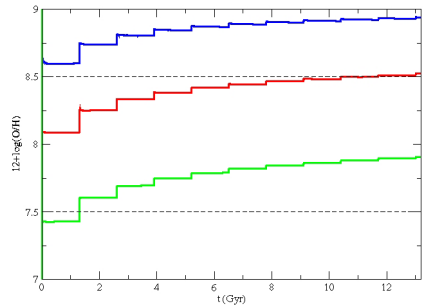}
%
%
\caption{Evolution of the oxygen abundance along 13.2 Gyrs M1(blue),
M2(red), M3(green).}
\label{fig:oh}       
\end{figure}

According to these results, the best models will be those with
efficiencies in the range $0.10 < \epsilon < 0.33$. From now on we
will analyze only these two types of models.
\subsection{Emission lines: diagnostic diagrams}
Once we have calculated the SED of the ionizing continuum, we have
applied the photo-ionization code to obtain the emission lines
intensities. Fig. 3 shows the [OIII]/H$\beta$  {\it vs} the
[OII]/H$\beta$ relation. Our models seem to reproduce the tendency
shown by data  \cite{hoy06,kna04}, with M2 following the high
excitation boundary of the data. However, the bulk of the data show
values of the [OII]/H$\beta$ ratio higher than computed. It should
however be taken into account that all our models end up with the
most recent burst. Therefore, from the point of view of gas
ionization, the model regions are unevolved.

\begin{figure}
\centering
\includegraphics[height=4cm]{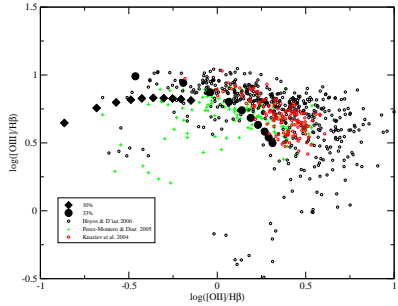}
%
%
\caption{Diagnostic diagrams comparing results from M2 and M3 with
observational data from \cite{hoy06,per05,kna04}}
\label{fig:diag}       
\end{figure}
\subsection{Equivalent width of H$\beta$}
In Fig.4a) we have represented the equivalent width of H$\beta$ {\sl
vs} the U-V color along the evolution of the galaxy for the
successive star bursts of M2. When a burst occurs the EW(H$\beta$)
is high and then decreases as the stellar population ages. We have
over-plotted the results for  SSPs from Starburst99 \cite{stb99}
with two different metallicities (left panel). In order to compare
our model results with observations, and since data on EW((H$\beta$)
{\sl vs} U-V color are scarce, we have used those from \cite{hoy06},
based in a pseudo-color computed from the intensities of the
adjacent continuum to the  [OII]$\lambda$3727 and
[OIII]$\lambda$5007 lines, producing a similar graph (right panel).
The trend of data is reproduced by our models, while this is not the
case for the SSP's. The black line in the left panel, that
represents a low metallicity SSP, does not reach the red U-V colors
shown by data nor the low values of EW((H$\beta$) observed for some
H{\sc ii} galaxies. In order to decrease EW(H$\beta$) and to obtain
bluer colors, a more metal-rich SSP might be selected (red line at
the left panel), but such high abundance is inconsistent with
observations, as we have already shown in Fig.2. Therefore, the
observed trend cannot be explained as either an age effect (low
metallicity SSPs do not reach EW(H$\beta$) values lower than 100\AA
and do not reproduce the colors), or a metallicity effect (high
metallicity SSPs are not observed in H{\sc ii} galaxies).  It is
necessary to include both effects simultaneously, as we have done,
to see how the effect of an underlying population, that contributes
to the color of the continuum making it redder than expected,
arises.
\begin{figure}
\centering
\includegraphics[height=4cm]{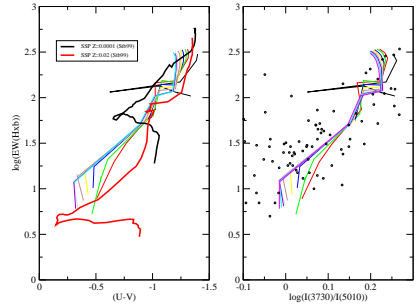}
%
%
\caption{{\it Left:}M2 equivalent width of the H$\beta$ line versus
(U-V) color, compared with SSPs from \cite{stb99}.{\it Right} Same,
but comparing with compare with observational data from
\cite{ter91}}
\label{fig:ewcont}       
\end{figure}
\section{Conclusions}
We have constructed models combining chemical evolution,
evolutionary synthesis and photoionization, in a self-consistent
way, to reproduce observational data of H{\sc ii} galaxies from
different works. Galaxies with 1 Kpc size, total mass of 10$^{8}$
M$\odot$ and a bursting star formation with different efficiencies
are modeled.
 Models with efficiencies higher than $\epsilon$=0.33 and lower than $\epsilon$=0.10 are not consistent with
 observed oxygen abundances and SFHs.
 The models reproduce the tendency of the observations in the [OIII]/$H\beta$ {\it vs } [OII]/H$\beta$ diagram
 following the high excitation boundary shown by data.
 The equivalent width of H$\beta$ line {\it vs}  U-V color shows that is
 necessary to include underlying populations in order  to reproduce the
 tendency of the observations, shifted to red colors, not only due
 to  a metallicity increase, but also to the contribution from the previous bursts
 continuum.
 Future work will include a complete grid of models with different
 attenuations and intermediate efficiencies to reproduce the whole
 set of H{\sc ii} galaxies data, and the application of these models
 to other kinds of galaxies, like luminous compact blue galaxies, which might be
 reproduced by the most efficient star formation models.

\printindex
\end{document}